\documentclass[pra,aps,eqsecnum,amsmath,amssymb,showpacs,showkeys]{revtex4}

\begin{document}
\clearpage
\preprint{}

\title{Partitioned trace distances}

\author{Alexey E. Rastegin}
\affiliation{Department of Theoretical Physics, Irkutsk State University,
Gagarin Bv. 20, Irkutsk 664003, Russia}
\email{rast@api.isu.ru}

\begin{abstract}
New quantum distance is introduced as a half-sum of several
singular values of difference between two density operators. This
is, up to factor, the metric induced by so-called Ky Fan norm. The
partitioned trace distances enjoy similar properties to the
standard trace distance, including the unitary invariance, the
strong convexity and the close relations to the classical
distances. The partitioned distances cannot increase under quantum
operations of certain kind including bistochastic maps. All the
basic properties are re-formulated as majorization relations.
Possible applications to quantum information processing are
briefly discussed.
\end{abstract}

\maketitle

\pagenumbering{arabic}
\setcounter{page}{1}

\section{Introduction}

A quantification of closeness of quantum states is inevitable task
for quantum information processing \cite{nielsen}. On the other
hand, the spaces of quantum states are very interesting
mathematical subjects \cite{bengtsson}. If states are pure then
comparison of them is not difficult. But all the real devices are
exposed to noise. So the pure states used will eventually evolve
to mixed states. This is not a unique reason for consideration of
mixed states. As it is shown in \cite{hfan03}, the cloning machine
which can use any mixed states in symmetric space is very
important in quantum computation. There are many ways to compare
two mixed quantum states. It has been found that two measures, the
trace distance \cite{nielsen,bengtsson} and the fidelity
\cite{uhlmann76,uhlmann83,jozsa94}, are widely useful in study of
quantum information. For instance, the fidelity function is most
frequently utilized as figure of merit for approximate  cloning
\cite{rast031,rast032}, data compression schemes \cite{szkola} and
quantum broadcasting \cite{holevo}. At the same time, the above
measures are not able to describe the problem of state closeness
in all respects. For example, the equality of fidelities for two
pairs of density operators does not imply their unitary
equivalence \cite{uhlmann00}. Recently, the sub-fidelity
\cite{uhlmann09} and the super-fidelity \cite{uhlmann09,foster}
have been studied. Some related measures were also used in the
literature, such as the Bures distance \cite{bengtsson}, the Monge
distance \cite{kzws01} and the sine distance \cite{rast06}. In
Ref. \cite{pati07,doherty} the Hilbert-Schmidt inner product was
utilized as figure of merit.

On the whole, reasons for use of some distance measure are mainly
provided by basic properties of the measure. These properties are
usually related to the measurements, change under quantum
operations and the convexity (concavity) in inputs. They must
ensure convenient mathematical formalism for study of processes in
quantum information. If some confidential information is encode in
quantum signals then any user will decode information by
measurements in the final stage. That is, after quantum
measurements he concludes from obtained data of measurement. The
authors of \cite{romano} gave the scheme in which obtained
statistics can then be used by the observer to reconstruct the
tested state without measurement back-action. In view of this, a
measure should be directly related to obtained data. Due to
numerous scenarios, we rather need some collection of reliable
measures complementing each other. As an example, the strength of
cryptosystem B92 with respect to state-dependent cloning is
revealed by relative error better than by global fidelity
\cite{rast02,rast033}. So the notion of relative error allows to
complete the picture of state-dependent cloning.

In the present work, we investigate a family of new distances
between mixed quantum states. These distances are closely related
to the trace distance which is obtained as particular case. Up to
a factor, each distance is metric induced by the Ky Fan norm. Like
the Schatten norms, the Ky Fan norms form a specially important
class of unitarily invariant norms. In general, the unitarily
invariant norms provide reasonable tools for obtaining distance
bounds on quantum information processing \cite{lidar1,lidar2}. The
distances induced by the Ky Fan norms enjoy the good features
similar to properties of the standard trace distance. By
construction, the described measures can naturally be called
``partitioned trace distances.'' In effect, together these
distances minutely characterize a distinguishability of two
quantum states via measurement statistics. It turns out that a set
of all the partitioned distances gives more detailed distinctions
of generated probability distributions than the standard trace
distance. We also obtain other properties which lighten use of the
partitioned trace distances.

\section{Definitions}

In this section, the definition of partitioned trace distances will be given. Let ${\cal{H}}$ be
$d$-dimensional Hilbert space. For any operator ${\mathsf{X}}$ on ${\cal{H}}$ the
operator ${\mathsf{X}}^{\dagger}{\mathsf{X}}$ is positive, that is
$\langle\psi|{\mathsf{X}}^{\dagger}{\mathsf{X}}|\psi\rangle\geq0$ for all $|\psi\rangle$.
The operator $|{\mathsf{X}}|$ is defined as unique positive square root of
${\mathsf{X}}^{\dagger}{\mathsf{X}}$. The eigenvalues of $|{\mathsf{X}}|$ counted with
multiplicities are called the singular values $s_j({\mathsf{X}})$ of operator
${\mathsf{X}}$ \cite{bhatia}. In the following, the singular values are arranged
in decreasing order, that is
$s_1({\mathsf{X}})\geq s_2({\mathsf{X}})\geq\cdots\geq s_d({\mathsf{X}})$.
For $k=1,\ldots,d$, the Ky Fan $k$-norm is defined as \cite{bhatia}
\begin{equation}
||{\mathsf{X}}||_{(k)}:=\sum\nolimits_{j=1}^{k}
s_j({\mathsf{X}}) \ .
\label{fannorm}
\end{equation}
The norm $||{\mathsf{X}}||_{(1)}$ is equal to the operator norm, and the norm
$||{\mathsf{X}}||_{(d)}={\rm tr}|{\mathsf{X}}|$
is the well-known trace norm \cite{bhatia}. Each operator norm induces some metric on quantum states.
In particular, the trace distance between quantum states $\rho$ and $\varrho$ is defined by \cite{nielsen}
\begin{equation}
D(\rho,\varrho):=\frac{1}{2}\>{\rm tr}|\rho-\varrho|\equiv\frac{1}{2}\>||\rho-\varrho||_{(d)}
\ . \label{tracedis}
\end{equation}
There is an alternative definition in terms of extremal properties of quantum operations \cite{rast07}.
The insertion of factor $1/2$ is justified by analogy with the classical distance and by the
bound $D(\rho,\varrho)\leq1$. Let $\{p_i\}$ and $\{q_i\}$ be two probability distributions over the same
index set. The $L_1$-distance (or Kolmogorov distance) is then defined as \cite{nielsen,bengtsson}
\begin{equation}
{\cal{D}}(p_i,q_i):=\frac{1}{2}\>\sum\nolimits_i |p_i-q_i| \ .
\label{cltracedis}
\end{equation}
In view of the above reasons, it is natural to introduce new distances in the following way.

{\bf Definition 2.1.} The $k$-th partitioned trace distance between density operator $\rho$ and
$\varrho$ is defined by
\begin{equation}
D_k(\rho,\varrho):=\frac{1}{2}\>||\rho-\varrho||_{(k)}
\ . \label{defpard}
\end{equation}

In simple case ${\rm dim}({\cal{H}})=2$, the difference
$\rho-\varrho=(1/2)({\vec{u}}-{\vec{v}})\cdot{\vec{\sigma}}$ has eigenvalues
$\pm(1/2)|{\vec{u}}-{\vec{v}}|$ in terms of corresponding Bloch vectors \cite{nielsen}. So we obtain
$D_1(\rho,\varrho)=(1/4)|{\vec{u}}-{\vec{v}}|$ and $D_2(\rho,\varrho)=(1/2)|{\vec{u}}-{\vec{v}}|$.
There are some clear properties of the introduced distances.
\begin{enumerate}
    \item{Bounds: $0\leq D_k(\rho,\varrho)\leq1$; $D_k(\rho,\varrho)=0$ if and only if $\rho=\varrho$.}
    \item{Symmetry: $D_k(\rho,\varrho)=D_k(\varrho,\rho)$.}
    \item{Triangle inequality: $D_k(\rho,\varrho)\leq D_k(\rho,\omega)+D_k(\omega,\varrho)$ for any three states $\rho$, $\omega$ and $\varrho$.}
    \item{If the states $\rho$ and $\varrho$ are pure then $D_1(\rho,\varrho)=(1/2)D_d(\rho,\varrho)$ and
$D_k(\rho,\varrho)=D_d(\rho,\varrho)$ for $k\geq2$.}
    \item{Unitary invariance: $D_k({\mathsf{U}}\rho{\mathsf{U}}^{\dagger},{\mathsf{U}}\varrho{\mathsf{U}}^{\dagger})=D_k(\rho,\varrho)$ for any unitary operator ${\mathsf{U}}$.}
\end{enumerate}

In the following, we will essentially use the statement known as Ky Fan's maximum principle \cite{bhatia}. Let
the eigenvalues $\lambda_j$ of Hermitian operator ${\mathsf{X}}$ be so arranged that
$\lambda_1\geq \lambda_2\geq\cdots\geq \lambda_d$. Then we have \cite{kyfan}
\begin{equation}
\sum\nolimits_{j=1}^{k} \lambda_j=
\max \{{\rm tr}({\mathsf{P}}{\mathsf{X}}):\>{\rm rank}({\mathsf{P}})=k\}
\ , \label{kfmaxpr1}
\end{equation}
where the maximization is over all projectors ${\mathsf{P}}$ of
rank $k$. Modifying the proof of theorem 1 of Ref. \cite{kyfan},
this principle can be re-formulated as
\begin{equation}
\sum\nolimits_{j=1}^{k} \lambda_j=
\max \{{\rm tr}({\mathsf{\Theta}}{\mathsf{X}}):
\>{\mathbf{0}}\leq{\mathsf{\Theta}}\leq{\mathbf{1}},\>{\rm tr}({\mathsf{\Theta}})=k\}
\ , \label{kfmaxpr2}
\end{equation}
where the maximum is taken over those positive operators ${\mathsf{\Theta}}$ with trace $k$ that satisfy
${\mathsf{\Theta}}\leq{\mathbf{1}}$. We do not enter into details here.

\section{Convexity properties}

In this section, some convexity properties of partitioned trace
distances will be established. It is known that extremal problems
are of great importance in applied disciplines. The presence of
convexity or concavity allows to simplify essentially a study of
many extremal problems \cite{rockaf}. Recall that Hermitian
operator $(\rho-\varrho)$ can be represented in the form
$\rho-\varrho={\mathsf{R}}-{\mathsf{T}}$, where ${\mathsf{R}}$ and
${\mathsf{T}}$ are positive operators with orthogonal support
spaces \cite{nielsen}. In linear algebra, this decomposition is
usually referred to as Jordan's decomposition \cite{bhatia}. Let
the $\varkappa_r$'s denote nonzero eigenvalues of ${\mathsf{R}}$,
and let the $\tau_t$'s denote nonzero eigenvalues of
${\mathsf{T}}$. By the spectral decomposition of $(\rho-\varrho)$,
we have
\begin{eqnarray}
{\mathsf{R}}&=\sum\nolimits_{r} \varkappa_r
\> |r\rangle\langle r| \ ,
\label{qeq} \\
{\mathsf{T}}&=\sum\nolimits_{t} \tau_t
\> |t\rangle\langle t| \ ,
\label{req}
\end{eqnarray}
where the eigenvectors are normalized. If we mutually rearrange the values $\varkappa_r$ and $\tau_t$ in
decreasing order then we obtain nonzero singular values $s_j$ of $(\rho-\varrho)$. It is clear that
$|\rho-\varrho|={\mathsf{R}}+{\mathsf{T}}$. By Ky Fan's maximum principle,
\begin{equation}
2D_k(\rho,\varrho)=
\max \{{\rm tr}({\mathsf{P}}|\rho-\varrho|):\>{\rm rank}({\mathsf{P}})\leq k\}
\ . \label{conv1}
\end{equation}
Here the condition ${\rm rank}({\mathsf{P}})\leq k$ is correct due to
$D_l(\rho,\varrho)\leq D_k(\rho,\varrho)$ for $l\leq k$.
Let us define the two specific subspaces for given $k$. The subspace ${\cal{L}}_R$ is spanned by those
$|r\rangle$'s that $\varkappa_r\in\{s_1,\ldots,s_k\}$. The subspace ${\cal{L}}_T$ is spanned by those
$|t\rangle$'s that $\tau_t\in\{s_1,\ldots,s_k\}$. The maximizing projector of minimal rank can
be written as ${\mathsf{P}}={\mathsf{P}}_R+{\mathsf{P}}_T$, where ${\mathsf{P}}_R$ is projector onto
${\cal{L}}_R$ and ${\mathsf{P}}_T$ is projector onto ${\cal{L}}_T$. If zeros are contained in the set
$\{s_1,\ldots,s_k\}$ then ${\rm rank}({\mathsf{P}})<k$. For the above projector we have
$({\mathsf{P}}_R-{\mathsf{P}}_T)(\rho-\varrho)={\mathsf{P}}_R{\mathsf{R}}+{\mathsf{P}}_T{\mathsf{T}}
\equiv{\mathsf{P}}|\rho-\varrho|$ and
\begin{equation}
2D_k(\rho,\varrho)=
{\rm tr}[({\mathsf{P}}_R-{\mathsf{P}}_T)(\rho-\varrho)]
\ . \label{conv3}
\end{equation}

{\bf Theorem 3.1.}
Let $\{p_i\}$ and $\{q_i\}$ be probability distributions over the same index set, and
$\rho_i$ and $\varrho_i$ be density operators with the same index. Then
\begin{equation}
D_k\left(\sum\nolimits_i p_i \rho_i,\sum\nolimits_i q_i \varrho_i\right)
\leq \sum\nolimits_i p_i D_k(\rho_i,\varrho_i) +{\cal{D}}(p_i,q_i)
\ . \label{theor3.1}
\end{equation}

{\bf Proof.} Let us put $\rho=\sum_i p_i \rho_i$ and $\varrho=\sum_i q_i \varrho_i$. Using
the Jordan decomposition of $(\rho_i-\varrho_i)$ and the triangle inequality for real numbers,
we see that
\begin{equation}
\left|{\rm tr}[({\mathsf{P}}_R-{\mathsf{P}}_T)(\rho_i-\varrho_i)]\right|
\leq    {\rm tr}({\mathsf{P}}|\rho_i-\varrho_i|)\ . \label{conv05}
\end{equation}
Further, $\left|{\rm tr}[({\mathsf{P}}_R-{\mathsf{P}}_T)\varrho_i]\right|\leq{\rm tr}\varrho_i=1$.
By these two inequalities and Eq. (\ref{conv3}), the doubled left-hand side of Eq. (\ref{theor3.1})
can be rewritten as
\begin{align}
&\sum\nolimits_i p_i {\rm tr}[({\mathsf{P}}_R-{\mathsf{P}}_T)(\rho_i-\varrho_i)]+
\nonumber\\
&\sum\nolimits_i (p_i-q_i) {\rm tr}[({\mathsf{P}}_R-{\mathsf{P}}_T)\varrho_i]
\nonumber\\
&\leq \sum\nolimits_i p_i {\rm tr}({\mathsf{P}}|\rho_i-\varrho_i|)
+\sum\nolimits_i |p_i-q_i|
\nonumber\\
&\leq \sum\nolimits_i 2p_iD_k(\rho_i,\varrho_i) +2{\cal{D}}(p_i,q_i)
\ , \label{conv5}
\end{align}
where the maximum principle has finally been used. $\square$

For the whole trace distance the proved property is called "strong
convexity" \cite{nielsen}. It must be stressed that the whole
classical distance ${\cal{D}}(p_i,q_i)$ is contained in Eq.
(\ref{theor3.1}). Indeed, the range of index $i$ is independent of
$k$. As a corollary of strong convexity, there is the joint
convexity. Namely,
\begin{equation}
D_k\left(\sum\nolimits_i p_i \rho_i,\sum\nolimits_i p_i \varrho_i\right)
\leq \sum\nolimits_i p_i D_k(\rho_i,\varrho_i)
\ . \label{conv6}
\end{equation}
Substituting $\varrho$ for all $\varrho_i$'s into Eq. (\ref{conv6}), we obtain the convexity in
the first input. That is,
\begin{equation}
D_k\left(\sum\nolimits_i p_i \rho_i, \varrho \right)
\leq \sum\nolimits_i p_i D_k(\rho_i,\varrho)
\ . \label{conv7}
\end{equation}
Due to symmetry we also have convexity in the second input
\cite{nielsen}. Using the reasons of this section, the triangle
inequality can easily be obtained. We refrain from presenting the
details here.

\section{Relations with the classicality}

Similar to the standard trace distance, the partitioned trace
distances can closely be related with the corresponding classical
distances. We shall now introduce the partitioned classical
distances between two probability distributions. By $n$ denote the
cardinality of distributions $\{p_i\}$ and $\{q_i\}$. For
$k=1,\ldots,n$, the $k$-th partitioned classical distances between
$\{p_i\}$ and $\{q_i\}$ is defined by
\begin{equation}
{\cal{D}}_k^{\downarrow}(p_i,q_i):=\frac{1}{2}\>\sum\nolimits_{i=1}^{k} |p_i-q_i|^{\downarrow}
\ , \label{pcltracedis}
\end{equation}
where the arrows down indicate that the absolute values are put in
the decreasing order. The distance ${\cal{D}}_n(p_i,q_i)$ is the
standard $L_1$-distance. [Because the distance
${\cal{D}}_n(p_i,q_i)$ contains all the differences $|p_i-q_i|$,
we justly omit the arrow down.] Let us suppose that two density
operators $\rho$ and $\varrho$ are commuting. So they are diagonal
in the same basis $\{|i\rangle\}$, that is
\begin{align}
& \rho=\sum\nolimits_i \mu_i
\>|i\rangle\langle i| \ , \label{rhocomm} \\
& \varrho=\sum\nolimits_i \nu_i
\>|i\rangle\langle i| \ . \label{varcomm}
\end{align}
It is clear that operator $\rho-\varrho=\sum\nolimits_i(\mu_i-\nu_i)|i\rangle\langle i|$ has singular
values $|\mu_i-\nu_i|^{\downarrow}$. Due to the definition of the partitioned trace distances and Eq.
(\ref{pcltracedis}), we have
\begin{equation}
D_k(\rho,\varrho)={\cal{D}}_k^{\downarrow}(\mu_i,\nu_i)
\ . \label{comm}
\end{equation}
Thus, if two density operators commute then the $k$-th partitioned trace distance between them is equal
to the $k$-th classical distance between their eigenvalues. A connection can also be posed in terms of
probabilities generated by a quantum measurement. A generalized measurement is described by so-called
"positive operator-valued measure" (POVM). Recall that POVM $\{{\mathsf{M}}_m\}$ is a set of positive
operators ${\mathsf{M}}_m$ satisfying \cite{nielsen,bengtsson}
\begin{equation}
\sum\nolimits_m {\mathsf{M}}_m={\mathbf{1}}
\ , \label{povmdef}
\end{equation}
where ${\mathbf{1}}$ is the identity operator. In general, this approach allows to extract more
information from a quantum system than the projective measurements. For the two density
operators, the traces ${\rm tr}({\mathsf{M}}_m\rho)\equiv p_m$ and ${\rm tr}({\mathsf{M}}_m\varrho)\equiv q_m$
are the probabilities of obtaining a measurement outcome labeled by $m$.

{\bf Theorem 4.1.}
For arbitrary two density operators $\rho$ and $\varrho$, there is
\begin{equation}
D_k(\rho,\varrho)=\max\{ {\cal{D}}_k^{\downarrow}(p_m,q_m):
\> {\rm{tr}}({\mathsf{M}}_m)\leq1 \}
\ , \label{relt0}
\end{equation}
where the maximum is taken over those POVMs that ${\rm{tr}}({\mathsf{M}}_m)\leq1$ for all the POVM elements.

{\bf Proof.} Using the expressions of probabilities $p_m$, $q_m$ and
$\rho-\varrho={\mathsf{R}}-{\mathsf{T}}$, we write
\begin{align}
2{\cal{D}}_k^{\downarrow}(p_m,q_m)&=\sum\nolimits_{m=1}^{k} |p_m-q_m|^{\downarrow}
\nonumber\\
&=\sum\nolimits_{m=1}^{k}
\left|{\rm tr}[{\mathsf{M}}_m({\mathsf{R}}-{\mathsf{T}})]\right|^{\downarrow}
\ . \label{relt1}
\end{align}
Due to ${\rm tr}({\mathsf{M}}_m{\mathsf{R}})\geq0$ and ${\rm tr}({\mathsf{M}}_m{\mathsf{T}})\geq0$, we
have
\begin{align}
\left|{\rm tr}[{\mathsf{M}}_m({\mathsf{R}}-{\mathsf{T}})]\right|
&\leq   {\rm tr}[{\mathsf{M}}_m({\mathsf{R}}+{\mathsf{T}})]\nonumber\\
&={\rm tr}({\mathsf{M}}_m|\rho-\varrho|)
\ . \label{relt2}
\end{align}
It follows from Eqs. (\ref{relt1}) and (\ref{relt2}) that
\begin{equation}
2{\cal{D}}_k^{\downarrow}(p_m,q_m)\leq
{\rm tr}({\mathsf{{\mathsf{\Theta}}}} |\rho-\varrho|)
\ , \label{relt3}
\end{equation}
where we put ${\mathsf{\Theta}}=\sum_{m=1}^{k}{\mathsf{M}}_m^{\>\downarrow}\>$. [Operators ${\mathsf{M}}_m^{\>\downarrow}$
are POVM elements rearranged with respect to the decreasing order of numbers $|p_m-q_m|^{\downarrow}$.]
Using the completeness relation (\ref{povmdef}) and ${\rm{tr}}({\mathsf{M}}_m)\leq1$, we get
${\mathsf{\Theta}}\leq{\mathbf{1}}$ and ${\rm tr}({\mathsf{\Theta}})\leq k$. Due to Eq.
(\ref{kfmaxpr2}), the right-hand side of Eq. (\ref{relt3}) is less than or equal to the sum of $k$ largest
eigenvalues of $|\rho-\varrho|$. In other words, $2{\cal{D}}_k^{\downarrow}(p_m,q_m)\leq2D_k(\rho,\varrho)$.
Let us show that the inequality is saturated for some POVM. We take the projector-valued measure
$\{|r\rangle\langle r|\}\cup\{|t\rangle\langle t|\}$, when each element
${\mathsf{M}}_m$ is either $|r\rangle\langle r|$ or $|t\rangle\langle t|$. For this measurement the value
$$
|p_m-q_m|=\left|{\rm tr}[{\mathsf{M}}_m({\mathsf{R}}-{\mathsf{T}})]\right|
$$
is equal to either $\varkappa_r$ or $\tau_t$. So the numbers $|p_m-q_m|^{\downarrow}$ are just
singular values of $(\rho-\varrho)$. We see that the equality in Eq. (\ref{relt0}) is reached by the
above PVM. $\square$

It should be stressed that Theorem 4.1 deals with POVMs whose
elements obey ${\rm{tr}}({\mathsf{M}}_m)\leq1$. At the same time,
for the whole trace distance the maximum can be taken over
arbitrary POVMs \cite{nielsen}. However, our restriction is hardly
essential from the operational point of view. By the Davies
theorem \cite{davies}, for many tasks the optimal POVM can be
built of elements of rank one. For such POVMs the statement of
Theorem 4.1 holds. Indeed, if ${\rm rank}({\mathsf{M}}_m)=1$ then
a single nonzero eigenvalue of element ${\mathsf{M}}_m$ does not
exceed one due to (\ref{povmdef}), whence
${\rm{tr}}({\mathsf{M}}_m)\leq1$. In the following, we will
discuss the result (\ref{relt0}) in the context of quantum
information processing.

\section{Formulation in terms of majorization}

In this section, we pose the above results within majorization
relations. Elements of majorization theory are fruitfully used in
the researches of quantum systems \cite{vidal}. For example, the
disorder criterion for separability has been found in terms of
majorization relations \cite{kempe}. We shall now recall the basic
notions of the theory of majorization \cite{olkin}. Let
$x=(x_1,\ldots,x_n)$ and $y=(y_1,\ldots,y_n)$ be elements of real
space ${\mathbb{R}}^n$. Let $x^{\downarrow}$ be the vector
obtained by rearranging the coordinates of $x$ in the decreasing
orders, that is
\begin{equation}
x_1^{\downarrow}\geq x_2^{\downarrow}\geq\cdots\geq x_n^{\downarrow}
\ . \label{xrear}
\end{equation}
We say that $x$ is {\it weakly submajorized} by $y$, in symbols $x\prec_w y$, when \cite{olkin}
\begin{equation}
\sum\nolimits_{j=1}^{k} x_j^{\downarrow} \leq
\sum\nolimits_{j=1}^{k} y_j^{\downarrow} \ ,\quad k=1,\ldots,n
\ . \label{major}
\end{equation}
If the inequality is saturated for $k=n$ then we say that $x$ is
{\it majorized} by $y$ \cite{bhatia,olkin}. In our analysis,
components of real vectors are positive. Let integer $n$ denote
the cardinality of set $\{{\mathsf{M}}_m\}$, so that $n$ is the
total number of  measurement outcomes. In the statement
(\ref{relt0}), the absolute values $|p_m-q_m|=|{\rm
tr}({\mathsf{M}}_m\rho)-{\rm tr}({\mathsf{M}}_m\varrho)|$ are
summands in ${\cal{D}}_k^{\downarrow}(p_m,q_m)$, the singular
values $s_j(\rho-\varrho)$ are summands in $D_k(\rho,\varrho)$.
Considering these summands as components of two real vectors, we
append zeros so that $|p-q|$ and $s(\rho-\varrho)$ have the same
dimension equal to $\max\{n,d\}$. Then the result (\ref{relt0})
can be re-formulated as follows.

{\bf Corollary 5.1.} If the condition
${\rm{tr}}({\mathsf{M}}_m)\leq1$ is fulfilled for all the POVM
elements then $|p-q|$ is weakly submajorized by $s(\rho-\varrho)$,
\begin{equation}
|p-q|   \prec_w s(\rho-\varrho)
\ . \label{corol51}
\end{equation}
If the equality ${\cal{D}}_n(p_m,q_m)=D_d(\rho,\varrho)$ is extra valid for POVM $\{{\mathsf{M}}_m\}$ then
\begin{equation}
|p-q|   \prec s(\rho-\varrho)
\ , \label{corol52}
\end{equation}
{\it id est} $|p-q|$ is majorized by $s(\rho-\varrho)$.

The relations (\ref{corol51}) and (\ref{corol52}) have the
advantage of physical interpretation of singular values
$s_j(\rho-\varrho)$ in terms of distinctions of probability
distributions. The relation between the standard trace distance
and the $L_1$-distance is the well-known result in quantum
information theory \cite{nielsen,bengtsson}. Namely,
\begin{equation}
D_d(\rho,\varrho)=\max\left\{ {\cal{D}}_n(p_m,q_m):
\>{\mathbf{0}}\leq{\mathsf{M}}_m\leq{\mathbf{1}},
\>\sum\nolimits_m{\mathsf{M}}_m={\mathbf{1}}\right\}
\ , \label{reltwh}
\end{equation}
where the maximum is taken over all the POVM measurements. So, if
two density operators are close in the standard trace distance,
then any measurement performed on those states will give
probability distributions close to each other. But they closeness
is still characterized by one quantity solely. In this regard, the
concept of partitioned trace distances provides sensitive and
flexible tools for comparing mixed quantum states. Instead of one
measure $D_d(\rho,\varrho)$, we now have a collection of $d$
measures $D_k(\rho,\varrho)$ and $d$ singular values
$s_j(\rho-\varrho)$. A closeness of two states are now described
by system of $d$ equations of the form (\ref{relt0}) or,
equivalently, by the majorization relation (\ref{corol51}).
Despite evident importance of the result (\ref{reltwh}), it does
not give as much detailed information as provided by the relation
(\ref{corol51}).

For each pair of states $\rho$ and $\varrho$ we have the specified
measurement such that the maximum in (\ref{relt0}) is reached for
all $k=1,\ldots,d$ (then $n=d$). As it follows from the proof of
Theorem 4.1, this measurement is described by the projector-valued
measure $\{|r\rangle\langle r|\}\cup\{|t\rangle\langle t|\}$. [The
vectors $|r\rangle$ and $|t\rangle$ are defined in (\ref{qeq}) and
(\ref{req}).] By simple calculation, we obtain
\begin{equation}
|p_j-q_j|^{\downarrow}=s_j(\rho-\varrho) \ , \quad k=1,\ldots,d
\ . \label{ent0}
\end{equation}
Thus, if the two probability distributions $\{p_j\}$ and $\{q_j\}$
are known then singular values $s_j(\rho-\varrho)$ may be estimate
due to (\ref{ent0}). Here we have a possibility of measurement of
partitioned trace distances in physical experiments. In principle,
this might be a subject of separate investigation.

\section{Monotonicity properties}

In this section we shall prove that the partitioned trace
distances cannot increase under quantum operation of certain kind.
The formalism of quantum operations provides a unified treatment
of possible state change in quantum theory \cite{nielsen}. Quantum
operations can be realized via programmable quantum processors
\cite{rosko}. Let ${\cal{H}}_A$ and ${\cal{H}}_B$ be
finite-dimensional Hilbert spaces. We will consider a map
${\cal{E}}$
\begin{equation}
\rho_A\to\rho_B:=
\frac{{\cal{E}}(\rho_A)}{{\rm{tr}}_B[{\cal{E}}(\rho_A)]}
\ , \label{oper1}
\end{equation}
where an input $\rho_A$ is normalized state on ${\cal{H}}_A$ and
an output $\rho_B$ is normalized state on ${\cal{H}}_B$. If the
map ${\cal{E}}$ describes physical process then it must be linear
and completely positive \cite{nielsen,bengtsson}. One demands that
$0\leq{\rm{tr}}_B[{\cal{E}}(\rho_A)]\leq1$ for each input $\rho_A$. Then the
map ${\cal{E}}$ is a quantum operation with the input space ${\cal{H}}_A$ and the
output space ${\cal{H}}_B$ \cite{nielsen,bengtsson}. Each completely positive map can be
written in the operator-sum representation. Namely, we have
\begin{equation}
{\cal{E}}(\rho_A)=\sum\nolimits_m {\mathsf{E}}_m
\,\rho_A\, {\mathsf{E}}_m^{\dagger}
\ , \label{oper3}
\end{equation}
where operators ${\mathsf{E}}_m$ map the input space ${\cal{H}}_A$ to
the output space ${\cal{H}}_B$ \cite{nielsen,bengtsson}. The normalization
implies that
\begin{equation}
\sum\nolimits_m {\mathsf{E}}_m^{\dagger}
{\mathsf{E}}_m \leq{\mathbf{1}}_A
\ , \label{oper4}
\end{equation}
where ${\mathbf{1}}_A$ is the identity on ${\cal{H}}_A$. When physical process
is deterministic, ${\rm{tr}}_B[{\cal{E}}(\rho_A)]=1$ and the upper bound in Eq. (\ref{oper4})
is saturated. Then a map is trace-preserving completely positive (TPCP). Such kind of operations
is quantum analogue of classical stochastic maps \cite{bengtsson}. A map ${\cal{E}}$ is unital when
\begin{equation}
{\cal{E}}({\mathbf{1}}_A)={\mathbf{1}}_B
\ , \label{oper5}
\end{equation}
where ${\mathbf{1}}_B$ is the identity on the output space ${\cal{H}}_B$. As it follows from Eqs.
(\ref{oper3}) and (\ref{oper5}), for unital operation
\begin{equation}
\sum\nolimits_m {\mathsf{E}}_m{\mathsf{E}}_m^{\dagger}
={\mathbf{1}}_B \ . \label{oper6}
\end{equation}
Of course, the conditions (\ref{oper4}) and (\ref{oper6}) are
independent from each other. For example, the depolarizing and
phase damping channels are unital, while amplitude damping is not
\cite{nielsen}. If TPCP-map is unital then it is called
"bistochastic" \cite{bengtsson}. This is quantum analogue of
bistochastic matrix, which is a stochastic matrix that leaves the
uniform probability vector invariant \cite{bengtsson}. Such
matrices can also be used for realization density matrix via
uniform ensemble \cite{ericsson}. Both the conditions
(\ref{oper4}) and (\ref{oper6}) are fulfilled for bistochastic
map.

{\bf Theorem 6.1.}
If TPCP-map satisfies the condition
\begin{equation}
\sum\nolimits_m {\mathsf{E}}_m {\mathsf{E}}_m^{\dagger}
\leq{\mathbf{1}}_B  \ , \label{teor0}
\end{equation}
then for arbitrary two normalized inputs $\rho_A$ and $\varrho_A$
\begin{equation}
D_k({\cal{E}}(\rho_A),{\cal{E}}(\varrho_A))\leq D_k(\rho_A,\varrho_A)
\ . \label{teor1}
\end{equation}

{\bf Proof.} In the following, $\rho_B\equiv{\cal{E}}(\rho_A)$ and
$\varrho_B\equiv{\cal{E}}(\varrho_A)$. As it has been shown in
Section 3, there exist two mutually orthogonal projectors
${\mathsf{\Pi}}_R$ and ${\mathsf{\Pi}}_T$ such that
\begin{align}
2D_k(\rho_B,\varrho_B)&=
{\rm tr}_B[({\mathsf{\Pi}}_R-{\mathsf{\Pi}}_T)(\rho_B-\varrho_B)]\nonumber\\
&={\rm tr}_A[({\mathsf{\Theta}}_R-{\mathsf{\Theta}}_T)(\rho_A-\varrho_A)]\nonumber\\
&\leq{\rm tr}_A[({\mathsf{\Theta}}_R+{\mathsf{\Theta}}_T)|\rho_A-\varrho_A|]
\ . \label{proof1}
\end{align}
Here we use the operator-sum representation, the properties of the trace and direct analogue of Eq.
(\ref{conv05}). We also put two positive operators
\begin{equation}
\left\{
\begin{array}{c}
{\mathsf{\Theta}}_R \\ {\mathsf{\Theta}}_T
\end{array}
\right\}
=\sum_m {\mathsf{E}}_m^{\dagger}
\left\{
\begin{array}{c}
{\mathsf{\Pi}}_R  \\ {\mathsf{\Pi}}_T
\end{array}
\right\}
{\mathsf{E}}_m \ . \label{thetar}\\
\end{equation}
Due to the precondition (\ref{teor0}) and ${\rm
rank}({\mathsf{\Pi}}_R+{\mathsf{\Pi}}_T)\leq k$, the trace of
positive operator
${\mathsf{\Theta}}={\mathsf{\Theta}}_R+{\mathsf{\Theta}}_T$
satisfies ${\rm tr}_A({\mathsf{\Theta}})\leq{\rm
tr}_B({\mathsf{\Pi}}_R+{\mathsf{\Pi}}_T)\leq k$. For any
$|\psi_A\rangle\in{\cal{H}}_A$ we have
\begin{align}
\langle\psi_A|{\mathsf{\Theta}}|\psi_A\rangle&=\sum\nolimits_m \langle\psi_A|{\mathsf{E}}_m^{\dagger}
({\mathsf{\Pi}}_R+{\mathsf{\Pi}}_T){\mathsf{E}}_m|\psi_A\rangle\nonumber\\
&\leq   \sum\nolimits_m \langle\psi_A|{\mathsf{E}}_m^{\dagger}
{\mathsf{E}}_m|\psi_A\rangle\leq\langle\psi_A|\psi_A\rangle
\label{oper7}
\end{align}
by $({\mathsf{\Pi}}_R+{\mathsf{\Pi}}_T)\leq{\mathbf{1}}_B$ and Eq. (\ref{oper4}). This implies that
${\mathsf{\Theta}}\leq{\mathbf{1}}_A$. Using Eq. (\ref{kfmaxpr2}), we then see
that the right-hand side of Eq. (\ref{proof1}) does not exceed
$2D_k(\rho_A,\varrho_A)$. $\square$

Thus, if TPCP-map satisfies the condition (\ref{teor0}) then it is
contractive with respect to all the partitioned trace distances.
In particular, a bistochastic map is contractive. If we allow
states to be not normalized then the condition of preservation of
the trace can be excluded. Due to Eq. (\ref{oper4}) the inequality
(\ref{oper7}) remains valid. So the studied distances cannot
increase under each quantum operation that obeys Eq.
(\ref{teor0}). The statement of Theorem 6.1 can be reformulated as
a majorization relation. Namely, if TPCP-map obeys the condition
(\ref{teor0}) then
\begin{equation}
s(\rho_B-\varrho_B)\prec_w  s(\rho_A-\varrho_A)
\ . \label{theoma}
\end{equation}
Together with the relation (\ref{corol51}), the majorization
relation (\ref{theoma}) allows to foresee some information on
measurement statistics at the output if statistics at the input of
quantum channel is known {\it a priori}.

It must be stressed that the whole trace distance cannot increase
under {\it arbitrary} trace-preserving quantum operation
\cite{nielsen}. For the partitioned trace distances this is not
the case. In effect, the Hilbert-Schmidt distance is also not
contractive generally \cite{wang}. Nevertheless, the class of
operations under which the partitioned distances are monotonous is
wide enough. This class contains both the basic transformations,
namely the unitary evolution and the measurement. In the case of
measurement, we have ${\cal{H}}_B={\cal{H}}_A$ and
\begin{equation}
\rho_B=\sum\nolimits_m{\mathsf{M}}_m^{1/2}\rho_A {\mathsf{M}}_m^{1/2}
\ . \label{mescas}
\end{equation}
Here the completeness relation implies both the trace preservation
and the unitality. Thus, no basic physical transformations ever
increase the partitioned trace distances.

\section{Conclusion}

To sum up, we see that the partitioned trace distances provide a
kind of physical interpretation for singular values of difference
between two density operator. Let us consider shortly possible
applications of the above results in quantum information
processing. First, many extensively used  channels satisfy the
condition (\ref{teor0}). In particular, this is fulfilled by
bistochastic maps, including the depolarizing channel and phase
damping channels. For these channels we can use all the
consequences of majorization relations (\ref{corol51}) and
(\ref{theoma}). Second, we may utilize the partitioned trace
distances in study of quantum channel given as ``black box.'' Here
we choose a set of probe states which is sufficiently dense in the
space of density operators. Putting these probe states into black
box, we may estimate the partitioned distances between outputs. If
an increase of some distance has been detected then the tested
quantum channel certainly violates the condition (\ref{teor0}). Of
course, this is rough draft of {\it modus operandi} only.

In the present paper we introduce a family of new distances
between mixed quantum states one of which is the trace distance.
In view of their construction, these distances have naturally be
called ``partitioned trace distances.'' Each distance is, up to
factor, a metric induced by some Ky Fan's norm. Several important
properties of introduced distances have been established. The
partitioned trace distances enjoy the metric properties, the
unitary invariance, the strong convexity and the close relations
to the corresponding classical distances. In addition, they do not
increase under quantum operations of certain kind including the
bistochastic maps. The partitioned trace distances provide
convenient tools for comparing mixed quantum states. So the
described family of distances can quite be utilized as figure of
merit for quantum information processing.

\acknowledgments

This work was supported in a part by the Ministry of Education and
Science of the Russian Federation under grants no. 2.2.1.1/1483,
2.1.1/1539. The present author is grateful to anonymous referees
for helpful remarks.

% Non-BibTeX users please use

\end{document}